\documentclass[12pt]{article}
\usepackage{graphicx}

\begin{document}
\title{Continuum effects on the pairing in neutron drip-line nuclei
studied with the canonical-basis HFB method}
\author{Naoki Tajima\\
Department of Applied Physics, Fukui University, 910-8507, Japan}
\date{\footnotesize Received: November 30, 2004 / Revised version: February 11, 2005}

\maketitle

\begin{abstract}
The canonical-basis HFB method provides an efficient way to describe
pairing correlations involving the continuum part of the single-particle
spectrum in coordinate-space representations.
It can be applied to super-conducting deformed drip-line nuclei
as easily as to stable or spherical nuclei.
This method is applied to a simulation of the approach to the
neutron drip line.  It turns out that
the HFB solution has a stronger pairing
and a smaller deformation as the Fermi level is raised.
However, such changes are smooth and finite.
No divergences or discontinuities of the radius or other quantities are found 
in the limit of zero Fermi energy.
The nuclear density continues to be localized even a little beyond 
the drip line.
\end{abstract}

\vspace*{-3mm}

\vspace*{6mm}

\baselineskip=0.6cm

In nuclei near the neutron drip line, the pairing correlation among
the neutrons involves significantly the continuum (positive-energy)
part of the Hartree-Fock (HF) single-particle states.  In principle,
there is no difficulty to treat such nuclei with the
Hartree-Fock-Bogoliubov (HFB) method, which is the framework to
incorporate the pairing correlation into mean-field approximations.
Indeed, there is no practical problem concerning spherical
nuclei \cite{DFT84,DNW96}.
However, deformed nuclei are not so easy to treat. 
The difficulty originates in the huge number of quasiparticle states,
most of which are spatially dislocalized continuum-spectrum states.
In the quasi-particle formalism, 
two methods have been used to overcome the difficulty,
one using transformed oscillator basis \cite{SDN03} and the other
using a coordinate mesh but only for axially symmetric nuclei \cite{TOU03}. 

Mathematically, HFB ground states can be expressed
in the form of the BCS variational function.
The single-particle states in this expression are localized.
They are called the HFB canonical basis.
This localization makes the level density of the canonical basis by far
smaller than that of the quasiparticle states because the former
is proportional to the volume of the nucleus while the latter to the
volume of the cavity to discretize the positive energy orbitals.
The canonical-basis HFB method enables one to obtain the canonical 
orbitals without knowing anything about the huge number of quasiparticle 
states. 
It can be applied to deformed neutron-rich nuclei without
difficulties.

The canonical-basis HFB method was originally introduced for spherical nuclei
in Ref.~\cite{RBR97}. 
I improved the method and implemented it 
for deformed nuclei \cite{Taj98a,Taj04}. 
I also found the necessity of momentum dependence for the contact
pairing interactions if one employs completely coordinate-space
representations (like three-dimensional Cartesian mesh, unlike the
radial mesh).

In quasiparticle HFB method, the canonical orbitals are obtained from the
one-body density matrix and thus 
people have not noticed the existence of a more direct 
relation to the Hamiltonian. The canonical-basis formalism discloses 
this relation. 
Namely, canonical orbitals above the Fermi level are roughly the bound
eigenstates of the pairing Hamiltonian.  It is not the HF Hamiltonian
which generates them.  This finding is helpful to understand the shell
structure in the continuum part of the spectrum \cite{Taj04}.


Now, let me show a result of a calculation performed with the 
canonical-basis method.
It is a simulation of the approach to the neutron drip line.
The system is the $N=Z=14$ nucleus.
Instead of increasing the difference $N-Z$, I modify
the parameters of the mean-field interaction.
Namely,  $t_0$ of the Skyrme force
is increased (toward zero from below) to raise the Fermi level while
$t_3 (>0)$ is decreased so that 
the saturation density of the  symmetric nuclear matter is unchanged.
This approach is taken only because the present version of my
computer program is designed for $N=Z$ systems.
I will examine the adequacy of this approach in future.

The interaction in the mean-field channel 
is the Skyrme SIII force \cite{BFG75} without the spin-orbit 
term. The coulomb force is also turned off.
Owing to the omission of these two interactions, the single-particle
states are four-fold degenerated.
I take into account 70 canonical basis states in each of the
four spin-isospin sectors.
The parameters of the pairing interaction \cite{Taj04} are
  $v_{\rm p}=-880$ MeV fm$^{3}$, 
  $k_{\rm c}=2$ fm$^{-1}$, and
  $\rho_{\rm c}=0.32$ fm$^{-3}$,
  and $\tilde{\rho}_{\rm c}=\infty$.

A three-dimensional Cartesian mesh representation is employed to
express the single-particle wavefunctions without assuming any spatial
symmetries. 
The mesh spacing is 0.8 fm while the edge of the cubic cavity is 40 fm. 

\begin{figure}
\begin{center}
\includegraphics[angle=0, width=15 cm]{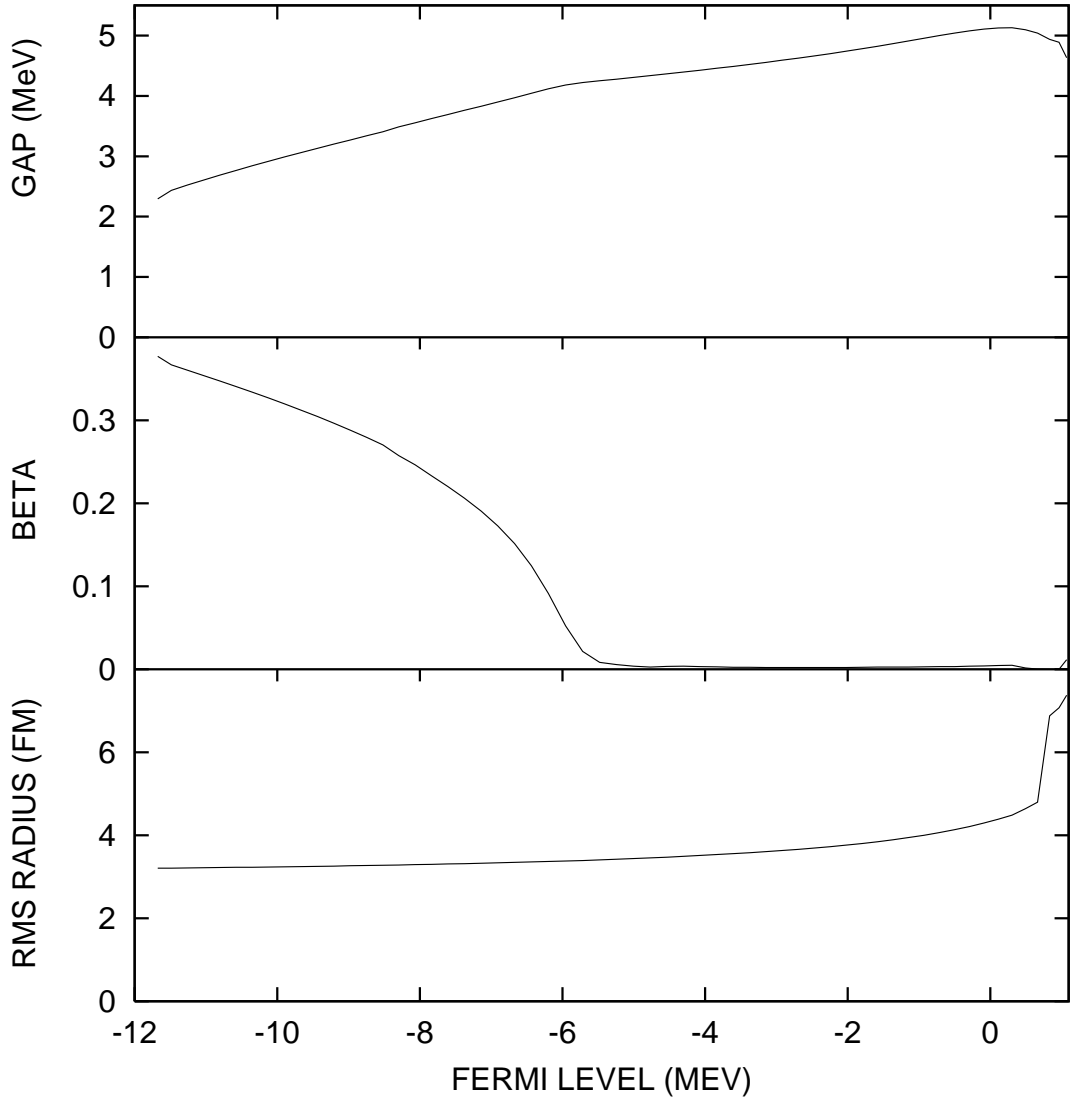}
\end{center}
\caption{
The average pairing gap (top), the quadrupole deformation parameter $\beta$
(middle), and the root-mean-square radius (bottom) plotted versus
the Fermi level.
}
\label{fig:1}
\end{figure}

Fig.~\ref{fig:1} shows how the HFB ground state changes
as the Fermi level ($\lambda$) rises.
The pairing gap is enhanced almost by a factor of two at the drip line
(where $\lambda = 0$ MeV) compared with the solution for the
original SIII force ($\lambda=-11.7$ MeV). 
The quadrupole deformation $\beta$
is decreased by the enhanced pairing.
The nucleus has a large prolate deformation at $\lambda=-11.7$ MeV but
becomes spherical for $\lambda > -5.5$ MeV.
On the other hand, the r.m.s.\ radius does not change so much.
Its increase is only 35 \% even at the drip line.
Dislocalization of the density occurs not at $\lambda=0$ MeV
but at higher $\lambda$ (0.8 MeV).
One can see only smooth changes at $\lambda \sim 0$ MeV.

\begin{figure}
\begin{center}
\includegraphics[angle=-90, width=15cm]{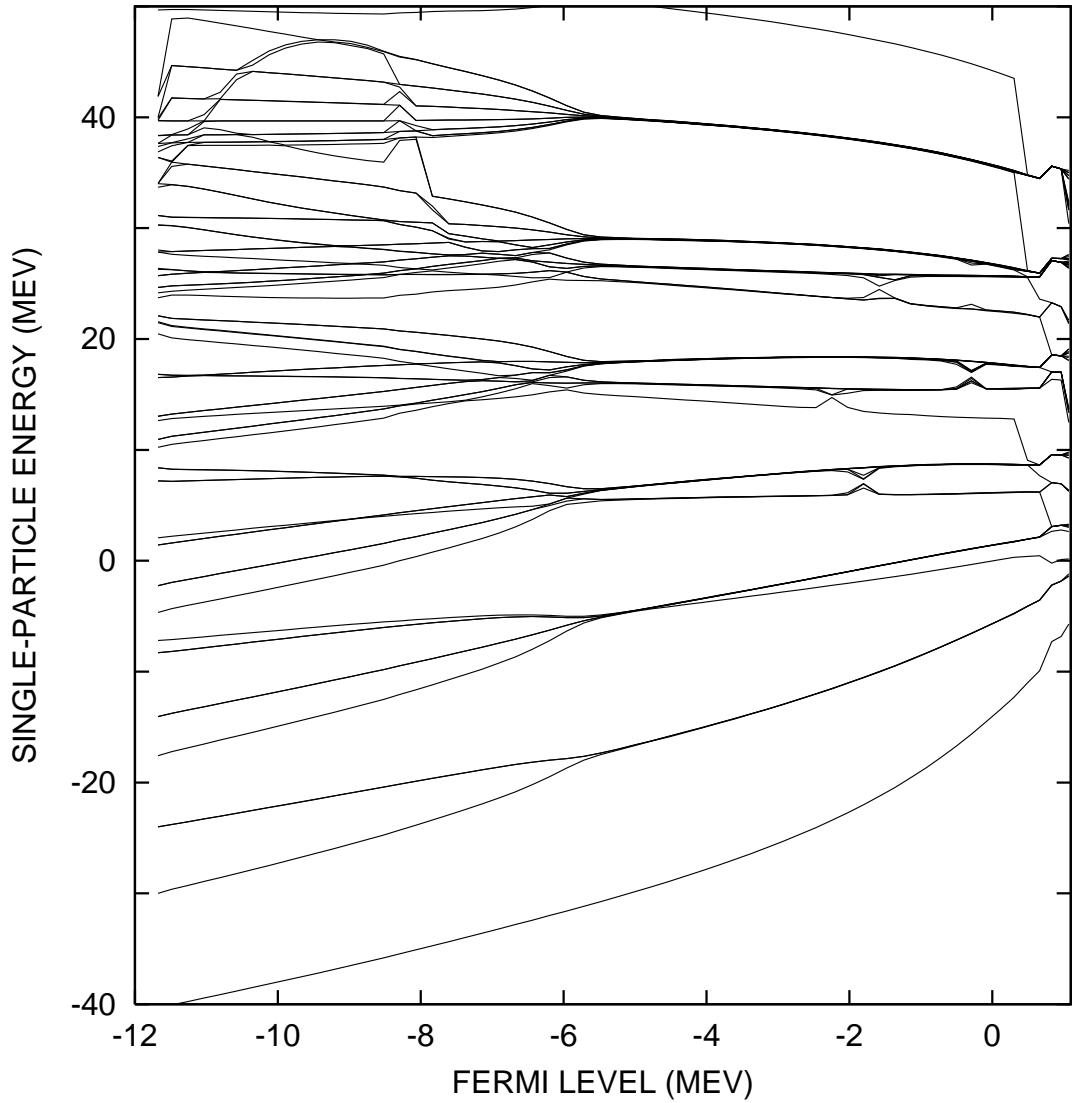}
\end{center}
\caption{
Expectation value of the HF Hamiltonian for each canonical-basis state
plotted versus the Fermi level.
}
\label{fig:2}
\end{figure}

Fig.~\ref{fig:2} shows the energies (expectation value of the
HF Hamiltonian) of canonical-basis states.
One can see that discrete bound states are obtained for both negative
and positive energies.
For $\lambda > -5.5$ MeV, the nucleus becomes spherical and the levels are
degenerated.
At $\lambda \sim -3$ MeV, the orbitals are (from the bottom)
s, p, s, d, p, f, s, d, g, etc.
Here again, there seems to be no violent changes at $\lambda \sim 0$ MeV.
Positive-energy localized s orbitals begin to spread over the cavity
only for $\lambda > $ 300 keV.

Considering general properties of HFB solutions\cite{DFT84}, 
the true ground state must be a dislocalized state.
The reason for the appearance of a localized solution seems to be as follows.
The dislocalization of an orbital requires
the increase of $v^2$ toward 1 because otherwise
the orbital is confined in the pairing potential, which is usually very deep
compared with the size of the kinetic energy term of the pairing Hamiltonian.
%
%
However, since $v^2=1$ corresponds to weaker pairing correlation and
larger total energy, a dislocalized solution is not necessarily reached 
in the iteration process of the gradient method
in certain circumstances like when the Fermi level is positive but low.
In contrast, with methods based on the diagonalization of the
quasi-particle Hamiltonian, a direct jump to $v^2=1$ can take place and
it results in the dislocalization of the corresponding orbital as soon
as it becomes energetically favorable.

The localized HFB solutions for positive Fermi energies obtained in the
canonical-basis formalism may be used as rough approximations to the
nuclei just beyond the neutron drip line.

\begin{center}
\framebox{
\begin{minipage}[t]{110mm}
This paper will be included in the Proceedings of the 4th International 
Conference on Exotic Nuclei and
Atomic Masses, September 12-16, 2004, Pine Mountain, Georgia, USA.
\end{minipage}
}
\end{center}


\begin{thebibliography}{1}

\bibitem{DFT84}
J.~Dobaczewski, H.~Flocard, and J.~Treiner,
Nucl.\ Phys. \textbf{A422}, 103 (1984)

\bibitem{DNW96}
J.~Dobaczewski, W.~Nazarewicz, T.~R. Werner, J.~F. Berger, C.~R. Chinn, and
  J.~Decharg{\'e},
Phys.\ Rev.\ C \textbf{53}, 2809 (1996)

\bibitem{SDN03}
M.~V.~Stoitsov, J.~Dobaczewski, W.~Nazarewicz, S.~Pittel, and D.~J.~Dean,
Phys.\ Rev.\ C \textbf{68}, 054312 (2003).

\bibitem{TOU03}
E.~Ter\'an, V.~E.~Oberacker, and A.~S.~Umar,
Phys.\ Rev.\ C \textbf{67}, 064314 (2003).

\bibitem{RBR97}
P.~G. Reinhard, M.~Bender, K.~Rutz, and J.~A. Maruhn,
Z.\ Phys. \textbf{A358}, 277 (1997)

\bibitem{Taj98a}
N.~Tajima,
in {\em Proceedings of the XVII RCNP International
  Symposium on Innovative Computational Methods in Nuclear Many-Body Problems,
Osaka, 1997,}
edited by H.~Horiuchi et al. (World Scientific, Singapore, 1998), p. 343

\bibitem{Taj04}
N.~Tajima, Phys.\ Rev.\ C \textbf{69}, 034305 (2004)

\bibitem{BFG75}
M.~Beiner, H.~Flocard, Nguyen van Giai, and P.~Quentin,
Nucl.\ Phys. \textbf{A238}, 29 (1975)

\end{thebibliography}
\end{document}